\documentclass[12pt]{article}

\usepackage[%
    geometry=yurie,
    caption=off,
    tikz=off,
    biblatex=yurie,
    hyperref=yurie
]{yurie}
\usepackage{yurie-math}
\usepackage{yurie-phy}
\usepackage{yurie-cft}

\numberwithin{equation}{section}

\allowdisplaybreaks

\newcommand{\deltaComplex}[1]{\delta_{\mathbb{C}}(#1)}
\newcommand{\deltaMC}{\delta^{(4)}}

\newcommand{\qhat}{\hat{q}}

\newcommand{\epsilonb}{\bar{\epsilon}}
\newcommand{\hwave}{\wave{h}}
\newcommand{\hbwave}{\wave{\bar{h}}}

\newcommand{\regionS}{I}

\newcommand{\threept}{C}

\newcommand{\ppb}{\bar{\partial}}

\newcommand{\threeptN}{c}

\newcommand{\graviton}{\mathsf{h}}
\newcommand{\gluon}{\mathsf{g}}

\newcommand{\inn}{\text{in}}
\newcommand{\out}{\text{out}}

\newcommand{\leading}{leading\xspace}

\begin{document}

\begin{titlepage}

    \title{Amplitude from crossing-symmetric celestial OPE}

    \author{Reiko Liu$^{a}$, Wen-Jie Ma$^{b,c}$}

    \date{}

    \maketitle\thispagestyle{empty}

    \address{${}^a$Yau Mathematical Sciences Center (YMSC), Tsinghua University, Beijing, 100084, China}

    \address{${}^b$Fudan Center for Mathematics and Interdisciplinary Study, Fudan University, Shanghai, 200433, China}

    \address{${}^c$Shanghai Institute for Mathematics and Interdisciplinary Sciences (SIMIS), Shanghai, 200433, China}

    \email{
        reiko@tsinghua.edu.cn,
        wenjie.ma@simis.cn
    }

    \vfill

    \begin{abstract}
        Assuming the existence of crossing symmetric celestial OPE, we propose a method to reconstruct four-point massless scattering amplitudes in the framework of celestial holography.
        This method relies only on CFT techniques and a remarkable property: scattering amplitudes can be derived from a single conformal block coefficient in celestial CFT.
        Utilizing this method, we reconstruct the MHV amplitudes in pure Yang-Mills, pure gravity, and Einstein-Yang-Mills theories.
    \end{abstract}

    \vfill

\end{titlepage}

%%%%%%%%%%%%%%%%%%%%%%%%%%%%%%%%%%%%%%%%%%%%%%%%%%%%%%%%%%%%%%%%

\begingroup
\hypersetup{linkcolor=black}
\tableofcontents
\thispagestyle{empty}
\endgroup

%%%%%%%%%%%%%%%%%%%%%%%%%%%%%%%%%%%%%%%%%%%%%%%%%%%%%%%%%%%%%%%%
\newpage
\setcounter{page}{1}

\section{Introduction}

Celestial holography establishes a connection between four-dimensional quantum gravity in asymptotically flat spacetime and a putative two-dimensional celestial conformal field theory (CCFT) \cite{Cheung:2016iub,Pasterski:2016qvg, Pasterski:2017kqt, Strominger:2017zoo, Raclariu:2021zjz, Pasterski:2021rjz, Pasterski:2021raf, McLoughlin:2022ljp}. Within this duality framework, the conformal correlators in the boundary CCFT, referred to as celestial amplitudes, are derived by expanding the scattering amplitudes using the conformal basis.

A major goal of celestial holography is to explore the scattering amplitudes by the techniques of conformal field theory (CFT). 
In CFT, conformal correlators can be expanded into conformal block (CB) coefficients through the CB expansion. 
Due to the existence of operator product expansion (OPE), the CB coefficient is further factorized into a product of an OPE coefficient and a three-point coefficient. 
Furthermore, since there are three different ways to implement the OPE, four-point correlators can be expanded into $\mathsf{s}$-, $\mathsf{t}$-, and $\mathsf{u}$-channel CB expansions.
This is known as the CFT crossing symmetry, one of the most powerful tools in CFT. 
Then the exchange operators and the corresponding CB coefficients are constrained by this crossing symmetry, serving as the foundation of modern nonperturbative conformal bootstrap program, see \eg the review \cite{Poland:2018epd}.

In a generic CFT, within the CB expansion of a single conformal correlator, the CB coefficients are typically independent. This makes it extremely challenging to solve the conformal bootstrap equations that arise from crossing symmetry. 
When focusing on CCFT, one may inquire whether the CB coefficients possess any distinctive properties to simplify the problem.
Research on CB expansions in CCFT can be found in \cite{Lam:2017ofc,Garcia-Sepulveda:2022lga,Atanasov:2021cje,Chang:2021wvv,Fan:2021pbp,Fan:2021isc,Fan:2022kpp,Chang:2022jut,Chang:2023ttm,Fan:2023lky,Himwich:2023njb,Liu:2024lbs,Surubaru:2025qhs}. 
It is worth noting that four-point massless celestial amplitudes do not admit an \schannel CB expansion due to momentum conservation. This issue can be resolved by using the shadow conformal basis and considering shadow celestial amplitudes \cite{Fan:2021isc,Fan:2021pbp,Chang:2022jut,Chang:2022seh,Liu:2024lbs,Surubaru:2025qhs}.

Surprisingly, in this paper, we discover that any massless four-point shadow celestial amplitude can be reconstructed from a particular CB coefficient, with all other coefficients being derivable from this one!
Based on this remarkable feature of CCFT, we develop a method of constructing scattering amplitudes.
The basic ingredients of our method are two CFT inputs: 1. existence of crossing symmetric OPE; 2. a particular OPE coefficient determined by translation/asymptotic symmetries \cite{Pate:2019lpp,Guevara:2021abz,Himwich:2021dau}.
This methodology paves the way for constructing scattering amplitudes using only CFT techniques.

\section{Background}

\textbf{(Shadow) celestial amplitude.}
The celestial amplitude $\cA^{h_i}(z_i)$ transforms as a conformal correlator with conformal weights $(h_i,\bar{h}_i)$ on the boundary celestial sphere.\footnote{In CFT, the conformal weights $(h,\hb)$ and the conformal dimension and spin $(\Delta,J)$ are related by $\Delta = h+\hb,\, J=h-\hb$. We will use the two notations interchangeably, and omit the $\zb$- and $\hb$-dependence if there is no ambiguity.}
For a $1^{a_1}+2^{a_2}\to 3^{a_3}+4^{a_4}$ scattering process of four massless particles with momenta $q_{i}$ and helicities $a_{i}$, the celestial amplitude $\mathcal{A}$ is related to the scattering amplitude $\cT$ by the Mellin transforms \cite{Pasterski:2016qvg,Pasterski:2017kqt}:
\begin{align}
    \label{eq:CA}
    &\cA^{h_i}(z_{i})
    =
    \bigg(\prod_{i=1}^{4} \int_0^{\infty}d\omega_i\,\omega_i^{\Delta_i-1}\bigg)
    \cT^{a_i}\bigl(q_i(\omega_{i},z_{i})\bigr)
    \, .
\end{align}
Here, the conformal spins are $J_{i}=a_{i}$ and the boundary coordinates $(z_{i},\zb_{i})$ are related to the on-shell momenta by
\begin{equation}
    q_{i}(\omega_{i},z_{i})=
    \omega_{i}\qhat_{i}(z_{i})
    =
    \omega_{i}
    (1+z_i \bar{z}_i,z_{i}+\bar{z}_{i},-i (z_{i}-\bar{z}_{i}),1-z_i \bar{z}_i)
    \, .
\end{equation}

Since scattering amplitudes are distributions of $\omega_{i},\, z_{i}$, the celestial amplitudes are distributions of $\Delta_{i},\, z_{i}$ accordingly.
In particular, due to momentum conservation, the celestial amplitude in \eqref{eq:CA} contains a factor $\delta(i\chi-i\bar{\chi})\theta(\chi-1)$ for $\chi=\frac{(z_{1}-z_{2})(z_{3}-z_{4})}{(z_{1}-z_{3})(z_{2}-z_{4})}$.
To remedy this distributional factor, the shadow celestial amplitude $\wave{\cA}$ is introduced by performing the shadow transform on the first particle
\begin{equation}
    \label{eq:ShadowCA}
    \wave\cA(z_{i})
    \eqq
    \intt{d^{2}z_{1'}}
    (z_{1}-z_{1'})^{2h_{1}-2}
    (\zb_{1}-\zb_{1'})^{2\bar{h}_{1}-2}
    \cA^{h_{i}}(z_{1'},\cdots)
    \, ,
\end{equation}
and the conformal weights of $\wave\cA$ are $(1-h_1,1-\hb_{1}), (h_i,\bar{h}_i)$ for $i=2,3,4$.

\section{Conformal block coefficients and scattering amplitudes}

In a generic CFT, determining a four-point correlator requires knowledge of all the exchange operators and their corresponding coefficients in the CB expansion.
However, for celestial CFT, we find that in the expansion of the shadow celestial amplitude $\wave{\cA}$, the coefficient associated with a particular exchange operator completely determines the scattering amplitude $\cT$ and consequently both $\cA$ and $\wave{\cA}$, irrespective of the OPE channels.
We refer to this particular exchange operator/CB coefficient as the \leading operator/coefficient and denote it as $\cO$/$\cC$.

To distinguish the OPE channels from the scattering channels, we only consider $1+2\to 3+4$ scatterings, and the labels $\schannel,\, \tchannel,\, \uchannel$ refer to the three OPE channels. To simplify the formulas, we adopt the multiple label notations for celestial sphere coordinates and conformal weights, \textit{e.g.} $\Delta_{a_{1}\cdots a_{n},b_{1}\cdots b_{m}}
\equiv
\sum_{i=1}^{n}\Delta_{a_{i}}-
\sum_{j=1}^{m}\Delta_{b_{j}}$.
We also introduce the following parameters for later convenience,
\begin{equation}
    \label{eq: abc}
    \alpha_{\schannel}\equiv 1+\half \Delta_{23,14}
    \, ,
    \quad
    \alpha_{\tchannel}\equiv 1+\half\Delta_{34,12}
    \, ,
    \quad
    \alpha_{\uchannel}\equiv 1+\half\Delta_{24,13}
    \, ,
    \quad
    \beta\equiv \Delta_{1234}-4
    \, .
\end{equation}

In the \schannel, the \leading operator $\cO^{\schannel}$ is the leading exchange with conformal weights $\Delta_{\schannel}=\Delta_{2,1}+2$, $J_{\schannel}=J_{2,1}$, and the \leading coefficient $\cC^{\schannel}_{\Delta_{\schannel},J_{\schannel}}$ is related to $\cT$ by
\begin{equation}
    \label{eq: C by scattering amplitude}
    \cC^{\schannel}_{\Delta_{\schannel},J_{\schannel}}(\alpha_{\schannel},\beta)
    =
    \inttt{dsdt}{\regionS}
    2^{-\beta-4}
    s^{\beta/2-\Delta_{\schannel}}
    (-t)^{\Delta_{\schannel}-\alpha_{\schannel}-1} (-u)^{\alpha_{\schannel}-1}
    \cT(s,t)
    \, .
\end{equation}
Here we have stripped off the kinematical part of the helicity amplitude as
\begin{equation}
    \label{eq: stripped scattering amplitude}
    \cT^{J_{i}}(q_{i})
    =
    \deltaMC(q_{1}+q_{2}-q_{3}-q_{4})
    H^{J_{i}}(z_{i})
    \cT(s,t)
    \, ,
\end{equation}
where $\cT(s,t)$ depends only on the Mandelstam variables $s=-(q_1+q_2)^2$, $t=-(q_1-q_3)^2$ and $u=-s-t$, and the explicit form of the kinematical factor $H$ can be found in the appendix.
The integration region is exactly the physical region of scattering
\begin{equation}
    \label{eq: s-channel physical region}
    \regionS\eqq
    \set{(s,t) \given t\leq 0\land s+t\geq 0}
    \, .
\end{equation}
After changing the variables from $(s,t)$ to $(\sqrt{s},-s/t)$, the integrals in \eqref{eq: C by scattering amplitude} become two independent Mellin transforms,
and the inverse is given by
\begin{align}
    \label{eq: scattering amplitude by C}
    \cT(s,t)
    &=
    \intt{[d\alpha_{\schannel}][d\beta]}
    2^{\beta +3}
    s^{-\beta /2}
    \Bigl(\frac{u}{t}\Bigr)^{-\alpha_{\schannel}}
    \Bigl(-\frac{s}{t}\Bigr)^{\Delta_{\schannel}}
    \cC^{\schannel}_{\Delta_{\schannel},J_{\schannel}}(\alpha_{\schannel},\beta)
    \, .
\end{align}
Here for each integral, the measure is $[d...]\eqq \frac{d...}{2\pi i}$, and the contour is a vertical line within the fundamental strip of the Mellin transform.

Similarly, in the $\tchannel$/\uchannel, the \leading operator has conformal dimension $\Delta_{\tchannel}=\Delta_{3,1}+2$ ($\Delta_{\uchannel}=\Delta_{4,1}+2$) and spin $J_{\tchannel}=J_{3,1}$ ($J_{\uchannel}=J_{4,1}$).
The relations between the scattering amplitude and the \leading coefficients $\cC^{\tchannel}_{\Delta_{\tchannel},J_{\tchannel}}$ and $\cC^{\uchannel}_{\Delta_{\uchannel},J_{\uchannel}}$ are
\begin{align}
    \label{eq: scattering amplitude by C T channel}
    &\theta(s+t)\theta(-t)\cT(s,t)
    =
    (-1)^{J_{1234}}
    \intt{[d\alpha_{\tchannel}][d\beta]}
    2^{\beta +3}
    (-t)^{-\beta /2}
    \Bigl(-\frac{s}{u}\Bigr)^{-\alpha_{\tchannel}}
    \Bigl(\frac{t}{u}\Bigr)^{\Delta_{\tchannel}}
    \cC^{\tchannel}_{\Delta_{\tchannel},J_{\tchannel}}
    (\alpha_{\tchannel},\beta)
    \, ,
    \\ \label{eq: scattering amplitude by C U channel}
    &\theta(s+t)\theta(-t)\cT(s,t)
    =
    (-1)^{J_{1234}}
    \intt{[d\alpha_{\uchannel}][d\beta]}
    2^{\beta +3}
    (-u)^{-\beta /2}
    \Bigl(-\frac{t}{s}\Bigr)^{-\alpha_{\uchannel}}
    \Bigl(-\frac{u}{s}\Bigr)^{\Delta_{\uchannel}}
    \cC^{\uchannel}_{\Delta_{\uchannel},J_{\uchannel}}
    (\alpha_{\uchannel},\beta)
    \, .
\end{align}

We stress here that it is the CFT crossing symmetry that allow us to construct $\cT(s,t)$ from three OPE channels, leading to \eqref{eq: scattering amplitude by C}, \eqref{eq: scattering amplitude by C T channel}, and \eqref{eq: scattering amplitude by C U channel}. It is also worthy noting that by \eqref{eq: C by scattering amplitude}, $\cC^{\schannel}$ depends only on the three combinations $(\Delta_{\schannel},\, \alpha_{\schannel},\, \beta)$ of the four external conformal dimensions $\Delta_i$, and by \eqref{eq: scattering amplitude by C}, $\Delta_{\schannel}$ can be freely chosen to obtain $\cT$.
Hence, only two out of four degrees of freedom are necessary. This is crucial for constructing scattering amplitudes later.

Moreover, we note that the $\tchannel/\uchannel$-channel is distinct from the \schannel.
In the future work \cite{celestialOPE}, we will show that only exchange operators of double-trace type appear in the \schannel OPE, with $\cO^{\schannel}$ being the leading exchange.
While in the $\tchannel$/$\uchannel$-channel, there can be other types of exchange operators, causing $\cO^{\tchannel}$/$\cO^{\uchannel}$ to be subleading.
Besides, due to the step functions in \eqref{eq: scattering amplitude by C T channel} and \eqref{eq: scattering amplitude by C U channel} compared with \eqref{eq: scattering amplitude by C}, from the $\tchannel/\uchannel$-channel OPE, we can only reconstruct the scattering amplitude inside the physical region.

\textbf{Analyticity and EFT.}
In \cite{Arkani-Hamed:2020gyp,Chang:2021wvv},
by the relation $\cT(s,t)\sim\intt{[d\beta]}s^{-\beta /2} \cA(\beta,-s/t)$ where $\cA(\beta,-s/t)$ is the celestial amplitude with conformal kinematical factors stripped off, the authors find that the $\beta$-analytic behaviors of $\cA$ manifest the UV/IR properties of the scattering amplitude $\cT$. More specifically, the residues of $\beta$-poles in the left half-plane are related to the Wilson coefficients in the EFT expansion, while the $\beta$-poles in the right half-plane arise from possible UV completions. Since \eqref{eq: scattering amplitude by C} is a two-fold Mellin transform of the \leading coefficient $\cC^{\schannel}$, the discussion can be naturally extended to the $(\alpha_{\schannel},\beta)$-planes. The $\beta$-analytic behaviors of $\cC^{\schannel}$ still reveal the UV/IR properties of $\cT$. By the relation to the scattering angle $\cot^2(\theta/2)=u/t$,
the forward (backward) scattering limit corresponds to the large (small) ratio $u/t$, hence is encoded in the residues of the $\alpha_{\schannel}$-poles in the right (left) half-plane.

\section{Reconstruction}

Since the CB coefficient is a product of OPE coefficient and three-point coefficient, \eqref{eq: scattering amplitude by C} implies that we can compute the scattering amplitude from a particular OPE coefficient and the associated three-point coefficient. 
The three-point coefficient is trivially fixed by the scaling behavior, \ie the action of bulk dilatation.
Moreover, certain OPE coefficients can be determined solely from translation/asymptotic symmetries, see \eg \cite{Pate:2019lpp,Guevara:2021abz,Himwich:2021dau}. 
Combining these two observations with \eqref{eq: scattering amplitude by C}, we can potentially construct scattering amplitudes without any detail of the bulk theory. In the following, we will present several examples to demonstrate the practicality of this method.

We first consider the MHV scattering of four gravitons $1^{--}+2^{--}\to 3^{++} +4^{++}$ in a pure gravity theory.
We denote an incoming/outgoing graviton operator with conformal dimension $\Delta$ and spin $\pm2$ by  $\graviton^{\inn/\out}_{\Delta,++/--}$.
In the \schannel, the OPE of outgoing gravitons is given by
\begin{align}
    \label{eq:GravitonGravitonOPE}
    \graviton^{\out}_{\Delta_3,++}(z_3)
    \graviton^{\out}_{\Delta_4,++}(z_4)
    \sim
    \frac{\bar{z}_{3,4}}{z_{3,4}}
    B(\Delta_3-1,\Delta_4-1)
    \graviton^{\out}_{\Delta_{34},++}(z_4)+\cdots
    \, .
\end{align}
Here $(...)$ denotes other possible primary operators, and the OPE coefficient $B(\Delta_3-1,\Delta_4-1)$ can be fixed either by translation/asymptotic symmetries under mild technical assumptions \cite{Pate:2019lpp,Guevara:2021abz,Himwich:2021dau}.
From \eqref{eq: C by scattering amplitude} the \leading operator $\cO_{\Delta_{2,1}+2,0}^{\schannel}$ is a scalar and hence the conformal spin does not match with \eqref{eq:GravitonGravitonOPE}.
In a pure gravity theory, we expect that $\cO^{\schannel}$ is built up from graviton operators, and one natural possibility is that $\cO^{\schannel}$ is a primary descendant in the conformal family of $\graviton^{\out}_{\Delta_{34},++}$.
As mentioned before, only two out of four $\Delta_i$-s are necessary for the reconstruction. Hence we can adjust $\Delta_{34}$ such that some descendant $\pp^{n}\ppb^{m}\graviton^{\out}_{\Delta_{34},++}$ becomes a primary scalar, and this fixes $n=0,\, m=2$ and $\Delta_{34}=1$.

If we assume the minimality condition: $\cO^{\schannel}_{\Delta_{2,1}+2,0}=\ppb^{2}\graviton^{\out}_{\Delta_{34},++}$, \ie there are no other primary scalars with dimension $\Delta_{34}$ in the OPE \eqref{eq:GravitonGravitonOPE}, then
\begin{equation}
    \label{eq:Graviton s-channel OPE matching}
    2-\Delta_1+\Delta_2=\Delta_3+\Delta_4+2=3
    \, ,
\end{equation}
and the \leading coefficient $\cC^{\schannel}$ is a product of OPE coefficient and three-point coefficient of $\bar{\partial}^2\graviton$.
The OPE coefficient can be read off from \eqref{eq:GravitonGravitonOPE} as $\half B(\Delta_3+1,\Delta_4-1)$.
From \eqref{eq:Graviton s-channel OPE matching} the three-point coefficient is associated with the celestial amplitude $\vev{
    \wave{\graviton}_{(2-\Delta_1,++)}^{\inn}
    \graviton_{(\Delta_1+1,--)}^{\inn}
    \LR{\bar{\partial}^2\graviton}_{(3,0)}^{\out}
}$.
Three-point celestial amplitudes of massless particles are not necessary vanishing, since the momentum conservation has solutions supported on collinear and soft regions \cite{Chang:2022seh}.
Moreover, by the scaling behavior, the three-point coefficient is fixed as $\half\threeptN_{\graviton}\deltaComplex{\Delta_2-1}$,\footnote{In the appendix, we explicitly compute some massless three-point celestial amplitudes and confirm this form.} where $\threeptN_{\graviton}$ is a numerical constant and $\delta_{\CC}$ is the complex $\delta$-function.\footnote{The complex $\delta$-function extends the Dirac $\delta$-function, see \eg \cite{Gelfand1,Gelfand2} and the applications in CCFT \cite{Donnay:2020guq,Liu:2024lbs}.}
Then the \leading coefficient can be written as
\begin{align}
    \label{eq:Cs Graviton}
    \cC^{\schannel}
    =
    \frac{1}{4}
    \threeptN_{\graviton}
    B(\Delta_3+1,\Delta_4-1)
    \deltaComplex{\Delta_2-1}
    \, .
\end{align}
By \eqref{eq:Cs Graviton} and \eqref{eq: scattering amplitude by C}, we obtain the stripped MHV amplitude of gravitons
\begin{align}
    \cT_{\graviton}(s,t)
    =
    \threeptN_{\graviton}
    \intt{[d\alpha_{\schannel}][d\beta]}
    2^{\beta +2}
    s^{-\beta /2}
    \Bigl(\frac{u}{t}\Bigr)^{-\alpha_{\schannel}}
    \Bigl(-\frac{s}{t}\Bigr)^{3}
    \deltaComplex{\beta+2}
    B(\alpha_s,1-\alpha_s)
    \, ,
\end{align}
where we have used $\Delta_{\schannel}=3$, $\beta=2\Delta_2-4$ and $\alpha_s=\Delta_3+1$ under \eqref{eq:Graviton s-channel OPE matching}.
The position of the $\alpha_{\schannel}$-contour is related to the asymptotic behavior of $\cT_{\graviton}$, and to distinguish different contour choices, we denote $n_{\schannel}+1\in\ZZ_{\geq 1}$ to be the smallest pole to the right of the contour. Then performing the $\alpha_{\schannel}$-integral we obtain
\begin{equation}\label{eq:cT Graviton s-channel}
    \cT_{\graviton}(s,t)
    =
    \threeptN_{\graviton}
    (-1)^{n_{\schannel}}s^{3} (-t)^{n_{\schannel}-2} (-u)^{-n_{\schannel}}
    \, .
\end{equation}

In the \tchannel OPE $\graviton^{\inn}_{\Delta_2,--}\xx\graviton^{\out}_{\Delta_4,++}$, there are four exchange gravitons \cite{Pate:2019lpp,Guevara:2021abz,Himwich:2021dau}, and it turns out that only two primary descendant with $\Delta_{24}=-3$ contribute, \textit{i.e.,}
\begin{align}
    \cO^{\tchannel}=c_1\partial^2\graviton^{\out}_{\Delta_{24},++}+c_2\partial^2\graviton^{\inn}_{\Delta_{24},++}
\end{align}
where $c_1$ and $c_2$ are numerical constants. The conformal dimensions are matched at
\begin{align}
    2-\Delta_1+\Delta_3=\Delta_2+\Delta_4+2=-1
    \, .
\end{align}
Following the same steps as the $\schannel$-channel case, we obtain
\begin{equation}\label{eq:cT Graviton t-channel}
    \cT_{\graviton}(s,t)
    =
    c_{2} s^{3+n_{\tchannel}} t^{-1}(-u)^{-1-n_{\tchannel}}
    -(c_{1}+c_{2})s^{2}t^{-1}
    \, ,
    \textInMath{for}
    n_{\tchannel}\in\NN
    \, .
\end{equation}
Comparing \eqref{eq:cT Graviton s-channel} with \eqref{eq:cT Graviton t-channel} in the physical region gives $n_{\schannel}=1$, $n_{\tchannel}=0$ and $\threeptN_{\graviton}=-c_{1}=c_{2}$.
Dropping the overall constant and restoring the little group factor, we reconstruct the graviton MHV amplitude as
\begin{equation}\label{eq:cT Graviton}
    \cT(1^{--} 2^{--} 3^{++} 4^{++})
    =
    \frac{z_{1,2}^2 \bar{z}_{3,4}^2}{z_{3,4}^2 \bar{z}_{1,2}^2}\frac{s^{3}}{tu}\delta^{(4)}(q_1+q_2-q_3-q_4)
    \, ,
\end{equation}

Using the same method, we have also successfully reconstructed the four gluon MHV amplitude in the pure Yang-Mills theory, as well as the following amplitude for three gluons and one graviton in the Einstein-Yang-Mills theory \cite{Stieberger:2016lng}
\begin{align}
    \cT^{abd}(1^{-}2^{-}3^{++}4^{+})
    =
    f^{abd}\frac{z_{1,2} z_{1,4}^{1/2} \bar{z}_{1,3}^{1/2} \bar{z}_{3,4}^{3/2}}{\bar{z}_{1,2} \bar{z}_{1,4}^{1/2} z_{1,3}^{1/2} z_{3,4}^{3/2}}
    \frac{s^{3/2}}{(-t)^{\frac{1}{2}}(-u)^{\frac{1}{2}}}\delta^{(4)}(q_1+q_2-q_3-q_4)
    \, .
\end{align}
Interested readers can refer to the appendix for details. It is worth noting that, due to the color factor dependence of gluons, in addition to the CFT crossing symmetry, we also utilized the permutation symmetry of identical particles when reconstructing the gluon MHV amplitudes.

\section{Discussion}

In this paper, we find that massless scattering amplitudes and consequently celestial amplitudes, can be derived from a particular OPE coefficient, as illustrated in \eqref{eq: scattering amplitude by C}, \eqref{eq: scattering amplitude by C T channel}, and \eqref{eq: scattering amplitude by C U channel}. 
This uncovers a fundamental difference in the structure of CCFT compared with generic CFTs.
Based on this novel structure and the CFT crossing symmetry, we develop a method to construct scattering amplitudes using only CFT techniques. 
Utilizing this method, we successfully reconstructed several amplitudes of gluons and gravitons.

Although the reconstructed scattering amplitudes are at tree-level, as the input OPE coefficient coincides with the tree-level collinear OPE \cite{Fan:2019emx,Pate:2019lpp}, our approach is nonperturbative.
Combining it with translation/asymptotic symmetries and current algebras \cite{He:2014laa,Kapec:2016jld,Strominger:2021mtt,Himwich:2021dau,Ball:2021tmb,Adamo:2021lrv,Himwich:2023njb,Banerjee:2023bni,Banerjee:2023zip,Pate:2019mfs,Agrawal:2024sju,Ball:2024oqa}, the celestial optical theorem \cite{Liu:2024vmx} related to bulk unitarity, and the crossing symmetry of scattering amplitudes, we aim to bootstrap nonperturbative gluon/graviton amplitudes.
A simple target model for bootstrap is the self-dual Yang-Mills/gravity theory, see particularly \cite{Costello:2022jpg,Costello:2022upu,Costello:2022wso}. We will pursue this as future work.

Currently our method is applicable only to four-point scattering amplitudes, it would be interesting to generalize it to higher-point amplitudes and, potentially, to other observables appearing at higher-points \cite{Caron-Huot:2023vxl,Borsten:2024dvq}. We expect to be able to reconstruct the higher-point MHV amplitudes.

Another direction is to extend our method to the celestial amplitudes and leaf amplitudes in Klein space. 
A recent proposal \cite{Melton:2024akx} suggests that a particular CFT, which combines the Liouville theory with free fermions, holographically produces MHV amplitudes in Klein space. If our method can be adapted, could one OPE coefficient determine the others in this dressed Liouville CFT?

Finally, to reconstruct the scattering amplitude, we assume that only particular primary descendants contribute to the \leading coefficient, named as the minimality condition.
The bulk interpretation of this assumption is obscure. For instance, is it related to the minimal coupling of bulk interactions, or perhaps with BCFW constructibility? 
Additionally, it would be intriguing to explore the physical meaning of the primary descendants, as their conformal dimensions are integers and might be associated with soft modes.

%%%%%%%%%%%%%%%%%%%%%%%%%%%%%%%%%%%%%%%%%%%%%%%%%%%%%%%%%%%%%%%%

\section*{Acknowledgements}

The authors would like to thank Jia-Qi Chen, Ban Lin and Zhi-Zhen Wang for useful discussions. 
WJM is supported by the National Natural Science Foundation of China No. 12405082 and Shanghai Magnolia Pujiang Talent Program A No. 24PJA118.

%%%%%%%%%%%%%%%%%%%%%%%%%%%%%%%%%%%%%%%%%%%%%%%%%%%%%%%%%%%%%%%%

\newpage
\appendix
\section*{Appendix}

In Appendix \ref{app: cT and cC} we derive the relation between the leading conformal block coefficient $\cC$ and the scattering amplitude $\cT$.
In Appendix \ref{app: three-point gluon} we compute the three-point coefficients $\threept$ of gluons.
In Appendix \ref{app: reconstruction} we provide the details of the \tchannel and the graviton reconstruction.

We adopt the following conventions and notations.
The bulk metric signature is $(-,+,+,+)$.
The integration measure on complex plane is $d^{2}z\eqq d\Re z d\Im z$.
The shadow conformal weights are denoted as $\wave{\Delta}\eqq 2-\Delta,\, \wave{J}\eqq -J$.
We also introduce the multiple label notations for coordinates $(z,\zb)$ and conformal weights $(\Delta,J),\, (h,\hb)$ as
\begin{align}
    \Delta_{a_{1}\cdots a_{n}}
    \equiv
    \sum_{i=1}^{n}\Delta_{a_{i}}
    \, ,
    \quad
    \Delta_{a_{1}\cdots a_{n},b_{1}\cdots b_{m}}
    \equiv
    \sum_{i=1}^{n}\Delta_{a_{i}}-
    \sum_{j=1}^{m}\Delta_{b_{j}}
    \, .
    \nn
\end{align}

\section{Leading conformal block coefficients}\label{app: cT and cC}

We first compute the four-point shadow celestial amplitude, then discuss its OPE limit in different channels and obtain the relation between $\cC$ and $\cT$, and finally provide an example of scalars.

We consider $1+2\to 3+4$ scatterings of four massless particles with helicities $J_{i}\in\ZZ$ and momenta $q_{i}$, parametrized by
\begin{equation}
    q_{i}=
    \omega_{i}\qhat_{i}
    =
    \omega_{i}
    (z_i \bar{z}_i+1,\bar{z}_{i}+z_{i},-i (z_{i}-\bar{z}_{i}),1-z_i \bar{z}_i)
    \, ,
\end{equation}
where $z_{i}\in\CC,\, \omega_{i}\geq 0$. The Mandelstam variables are $s=-(q_{1}+q_{2})^{2}$, $t=-(q_{1}-q_{3})^{2}$ and $u=-s-t$.
By the little group rescaling, the helicity amplitude $\cT$ can be factorized into
\begin{equation}
    \label{eqapp: stripped scattering amplitude}
    \cT^{J_{i}}(q_{i})
    =
    \deltaMC(q_{1}+q_{2}-q_{3}-q_{4})
    H^{J_{i}}(z_{i})
    \cT(s,t)
    \, .
\end{equation}
Here we choose the polarization vectors to be $\epsilon_{i}=\pp_{z_{i}}\qhat_{i}$ for $J_{i}=1$ and $\epsilonb_{i}=\pp_{\zb_{i}}\qhat_{i}$ for $J_{i}=-1$, so that the spinor products are $\braketS{ij}=2\sqrt{\omega_i\omega_j}\,  \bar{z}_{i,j}$ and $\braketA{ij}=-2\sqrt{\omega_i\omega_j}\,  z_{i,j}$, and the little group factor can be chosen as
\begin{equation}
    \label{eqapp: cT kinematical factor}
    H^{J_{i}}(z_{i})
    =
    z_{1,2}^{-\frac{J_{12}}{2}}
    z_{1,3}^{\frac{J_{4,3}}{2}}
    z_{1,4}^{\frac{J_{23,14}}{2}}
    z_{2,4}^{\frac{J_{1,2}}{2}}
    z_{3,4}^{-\frac{J_{34}}{2}}
    \times
    (z_{i}\to \zb_{i},J_{i}\to -J_{i})
    \, .
\end{equation}

In the CFT side, the four-point correlator $\cA$ can be factorized into the stripped one together with a kinematical factor
\begin{equation}
    \cA^{h_{i}}(z_{i})
    =
    \COK^{{h_{i}}}(z_{i})
    \cA^{h_{i}}(\chi)
    \, ,
\end{equation}
where $\chi=\frac{z_{1,2} z_{3,4}}{z_{1,3} z_{2,4}}$.
For the \schannel OPE, the conventional choice is
\begin{equation}
    \label{eqapp: kinematical factor}
    \COK^{{h_{i}}}(z_{i})
    =
    z_{1,2}^{-h_{12}}
    z_{1,3}^{h_{4,3}}
    z_{1,4}^{h_{23,14}}
    z_{2,4}^{h_{1,2}}
    z_{3,4}^{-h_{34}}
    \times
    (z_{i}\to \zb_{i},h_{i}\to \hb_{i})
    \, .
\end{equation}
Then an exchange primary operator $\cO_{(h_{0},\hb_{0})}$ in this channel contributes to
\begin{equation}
    \cA^{h_{i}}(\chi)\sim \cC^{h_{i}}_{h_{0},\hb_{0}}\chi^{h_{0}}\chib^{\hb_{0}}
    \, .
\end{equation}
Here $\cC$ is called the \leading coefficient associated with $\cO_{\Delta_{0},J_{0}}$ and can be factorized into a product of OPE coefficient and three-point coefficient.

\subsection{Shadow celestial amplitudes}

To be motivated, we start from the case of four scalars.
For a scalar scattering amplitude $\cT(q_{1},\cdots)$, the shadow transform \wrt the first particle can be recast into
\begin{align}
    \wave\cA(z_{1},\cdots)
    &=
    \intt{d^{2}z_{1'}}
    |z_{1,1'}|^{2\Delta_1-4}
    \intrange{d\omega_{1}}{0}{\oo}
    \omega_{1}^{\Delta_1-1}
    \cT(q_{1'}(\omega_{1},z_{1'}),\cdots)
    \\
    &=
    \intt{d^{4}q_{1'}}
    \delta(q_{1'}^{2})
    2^{1-\Delta_1}
    \bigl(-q_{1'} \cdot \hat{q}_1(z_{1})\bigr)^{\Delta_1-2}
    \cT(q_{1'},\cdots)
    \nn
    \, .
\end{align}
Then the shadow celestial amplitude is
\begin{align}
    \wave\cA
    &=
    \intrange{d\omega_{2}}{0}{\oo}
    \omega_2^{\Delta_2-1}
    \intrange{d\omega_{3}}{0}{\oo}
    \omega_3^{\Delta_3-1}
    \intrange{d\omega_{4}}{0}{\oo}
    \omega_4^{\Delta_4-1}
    \int d^{4}q_{1'}\delta(q_{1'}^{2})
    \cT(s,t)
    \\
    &\peq \xx
    2^{1-\Delta_1}
    (-q_{1'} \cdot \hat{q}_1)^{\Delta_1-2}
    \deltaMC(q_2-q_3-q_4+q_{1'})
    \nn
    \, .
\end{align}
Writing the Mandelstam variables $(s,t)$ into momenta, we can perform the $q_{1'}$-integral by the momentum conservation,
\begin{align}
    \wave\cA
    &=
    \intrange{d\omega_{2}}{0}{\oo}
    \omega_2^{\Delta_2-1}
    \intrange{d\omega_{3}}{0}{\oo}
    \omega_3^{\Delta_3-1}
    \intrange{d\omega_{4}}{0}{\oo}
    \omega_4^{\Delta_4-1}
    \cT(-2 \omega_3 \omega_4 \hat{q}_{34},2 \omega_2 \omega_4 \hat{q}_{24})
    \\
    &\peq \xx
    2^{1-\Delta_1}
    (\hat{q}_{12} \omega_2-\hat{q}_{13} \omega_3-\hat{q}_{14} \omega_4)^{\Delta_1-2}
    \delta (
        -2 \hat{q}_{23} \omega_2 \omega_3
        -2 \hat{q}_{24} \omega_2 \omega_4
        +2 \hat{q}_{34} \omega_3 \omega_4
    )
    \nn
    \, ,
\end{align}
where $\qhat_{ij}\eqq \qhat_{i}\cdot\qhat_{j}=-2 \abs{z_{i,j}}^{2}$.
Then changing $(\omega_2,\omega_3)$ back to
\begin{equation}
    \label{eq: st to omega}
    s=-2 \omega_3 \omega_4 \hat{q}_{34}>0
    \, ,
    \quad
    t=2 \omega_2 \omega_4 \hat{q}_{24}<0
    \, ,
\end{equation}
we obtain
\begin{align}
    \wave\cA
    &=
    \inttt{dsdt}{\regionS}
    \intrange{d\omega_{4}}{0}{\oo}
    2^{3-2 \Delta_1-\Delta_{23}}
    s^{\Delta_3-1}
    (-t)^{\Delta_2-1}
    \omega_4^{1+\Delta_{4,123}}
    \cT(s,t)
    \\
    &\peq\xx
    (-\hat{q}_{24})^{-\Delta_2}
    (-\hat{q}_{34})^{-\Delta_3}
    \bigg(
        \frac{t \hat{q}_{12}}{\hat{q}_{24}}+\frac{s \hat{q}_{13}}{\hat{q}_{34}}
        -2 \hat{q}_{14} \omega_4^2
    \bigg)^{\Delta_1-2}
    \delta\bigg(
        u+\frac{s t \hat{q}_{23}}{2 \hat{q}_{24} \hat{q}_{34} \omega_4^2}
    \bigg)
    \nn
    \, .
\end{align}
Notice that the integration region of $(s,t)$ has shrunk from \eqref{eq: st to omega} to the physical region of $1+2\to 3+4$ scattering,
\begin{equation}
    \label{eqapp: s-channel physical region}
    \regionS
    =
    \set{(s,t) \given t\leq 0\land s+t\geq 0}
    \, .
\end{equation}
The reason is that, for any $\qhat_{2},\, \qhat_{3},\, \qhat_{4}$ and $\omega_4 \geq 0$, the $\delta$-function has a solution at
\begin{equation}
    \omega_{4}^{2}=
    -\frac{\hat{q}_{23} s t}{2 \hat{q}_{24} \hat{q}_{34} u}
    \, ,
\end{equation}
if and only if the Mandelstam variables $(s,t)$ belong to the physical region $\regionS$.
Then we can perform the $\omega_{4}$-integral, leading to
\begin{align}
    \label{eq: shadow amplitude derivation step A}
    \wave\cA
    &=
    \inttt{dsdt}{\regionS}
    2^{\frac{2-3 \Delta_1-\Delta_{234}}{2}}
    s^{\frac{\Delta_{34,12}}{2}}
    (-t)^{\frac{\Delta_{24,13}}{2}}
    (-u)^{\frac{\Delta_{23,14}}{2}}
    \cT(s,t)
    \\
    &\peq\xx
    (-\hat{q}_{23})^{\frac{2+\Delta_{4,123}}{2}}
    (-\hat{q}_{24})^{\frac{2+\Delta_{3,124}}{2}}
    (-\hat{q}_{34})^{\frac{2+\Delta_{2,134}}{2}}
    (-s t \hat{q}_{14} \hat{q}_{23}-s u \hat{q}_{13} \hat{q}_{24}-t u \hat{q}_{12} \hat{q}_{34})^{\Delta_1-2}
    \, .
    \nn
\end{align}
The $\qhat_{ij}$-dependencies can be recast into cross ratios $\chi_{ijkl}=\frac{\qhat_{ij}\qhat_{kl}}{\qhat_{ik}\qhat_{jl}}$ as
\begin{align}
    \wave\cA(z_{i})
    &=
    \COK^{{h_{i}}}(z_{i})
    \inttt{dsdt}{\regionS}
    2^{-\Delta_{1234}}
    s^{\frac{\Delta_{34,12}}{2}}
    (-t)^{\frac{\Delta_{24,13}}{2}}
    (-u)^{\frac{\Delta_{23,14}}{2}}
    \cT(s,t)
    \\
    &\peq\xx
    \chi_{2134}^{\frac{\Delta_{12}-2}{2}}
    \chi_{3124}^{\frac{\Delta_{3,4}}{2}}
    (
        -t u
        -s u \chi_{1324}
        -s t \chi_{2314}
    )^{\Delta_1-2}
    \, ,
    \nn
\end{align}
where the kinematical factor $\COK$ is given in \eqref{eqapp: kinematical factor}.
Then fixing to the conformal frame, the stripped shadow celestial amplitude is
\begin{align}
    \wave{\cA}(\chi)
    &=
    2^{-\Delta_{1234}}
    \chi^{1+\frac{\Delta_{2,1}}{2}}
    \bar{\chi}^{1+\frac{\Delta_{2,1}}{2}}
    (1-\chi )^{1+\frac{\Delta_{4,123}}{2}}
    (1-\bar{\chi})^{1+\frac{\Delta_{4,123}}{2}}
    \\
    &\peq\xx
    \inttt{dsdt}{\regionS}
    s^{\frac{\Delta_{34,12}}{2}}
    (-t)^{\frac{\Delta_{24,13}}{2}}
    (-u)^{\frac{\Delta_{23,14}}{2}}
    (s+t \chi )^{\Delta_1-2}
    (s+t \bar{\chi})^{\Delta_1-2}
    \cT(s,t)
    \, .
    \nn
\end{align}

For the helicity scattering amplitude \eqref{eqapp: stripped scattering amplitude}, there is only one essential difference: to perform the $q_{1'}$-integral, we need to rewrite the spinning shadow transform as well as the little group factor in \eqref{eqapp: cT kinematical factor} into covariant forms. This can be achieved by $q_i \cdot \epsilon_j = 2 \omega_i \bar{z}_{i,j}$ and $q_i \cdot \bar{\epsilon}_j = 2 \omega_i z_{i,j}$. For the spinning shadow transform, we have
\begin{align}
    &\peq
    \intt{d^{2}z_{1'}}
    z_{1,1'}^{2 h_1-2}
    \zb_{1,1'}^{2 \hb_1-2}
    \intrange{d\omega_{1}}{0}{\oo}
    \omega_{1}^{\Delta_1-1}
    \cdots
    \\
    &=
    \intt{d^{4}q_{1'}}
    \delta(q_{1'}^{2})
    2^{1-\Delta_1}
    \bigl(-q_{1'} \cdot \hat{q}_1(z_{1})\bigr)^{\Delta_1-2}
    \biggl(\frac{q_{1'} \cdot \epsilon_1(z_{1})}{q_{1'} \cdot \bar{\epsilon}_1(z_{1})}\biggr)^{-J_1}
    \cdots
    \nn
    \, ,
\end{align}
and similarly for the little group factor, we have
\begin{equation}
    H^{J_{i}}
    =
    \biggl(\frac{q_{1'} \cdot \bar{\epsilon}_2}{q_{1'} \cdot \epsilon_2}\biggr)^{-\frac{J_{12}}{2}}
    \biggl(\frac{q_{1'} \cdot \bar{\epsilon}_3}{q_{1'} \cdot \epsilon_3}\biggr)^{\frac{J_{4,3}}{2}}
    \biggl(\frac{q_{1'} \cdot \bar{\epsilon}_4}{q_{1'} \cdot \epsilon_4}\biggr)^{\frac{J_{23,14}}{2}}
    \biggl(\frac{q_2 \cdot \bar{\epsilon}_4}{q_2 \cdot \epsilon_4}\biggr)^{\frac{J_{1,2}}{2}}
    \biggl(\frac{q_3 \cdot \bar{\epsilon}_4}{q_3 \cdot \epsilon_4}\biggr)^{-\frac{J_{34}}{2}}
    \, .
\end{equation}
The remaining steps are the same as the scalar case. For instance, the equation \eqref{eq: shadow amplitude derivation step A} now acquires additional factors
\begin{align}
    &
    \biggl(
        \frac{
            t u \hat{q}_2 \cdot \epsilon_1 \hat{q}_{34}+s u \hat{q}_{24} \hat{q}_3 \cdot \epsilon_1+s t \hat{q}_{23} \hat{q}_4 \cdot \epsilon_1
        }{
            t u \hat{q}_2 \cdot \bar{\epsilon}_1 \hat{q}_{34}+s u \hat{q}_{24} \hat{q}_3 \cdot \bar{\epsilon}_1+s t \hat{q}_{23} \hat{q}_4 \cdot \bar{\epsilon}_1
        }
    \biggr)^{-J_1}
    \biggl(
        \frac{\qhat_2 \cdot \epsilon_4}{\qhat_2 \cdot \bar{\epsilon}_4}
    \biggr)^{-\frac{J_{1,2}}{2}}
    \biggl(
        \frac{\qhat_3 \cdot \epsilon_4}{\qhat_3 \cdot \bar{\epsilon}_4}
    \biggr)^{\frac{J_{34}}{2}}
    \xx
    \\
    &
    \biggl(
        \frac{
            u \hat{q}_{24} \hat{q}_3 \cdot \epsilon_2+t \hat{q}_{23} \hat{q}_4 \cdot \epsilon_2
        }{
            u \hat{q}_{24} \hat{q}_3 \cdot \bar{\epsilon}_2+t \hat{q}_{23} \hat{q}_4 \cdot \bar{\epsilon}_2
        }
    \biggr)^{\frac{J_{12}}{2}}
    \biggl(
        \frac{
            t \hat{q}_2 \cdot \epsilon_4 \hat{q}_{34}+s \hat{q}_{24} \hat{q}_3 \cdot \epsilon_4
        }{
            t \hat{q}_2 \cdot \bar{\epsilon}_4 \hat{q}_{34}+s \hat{q}_{24} \hat{q}_3 \cdot \bar{\epsilon}_4
        }
    \biggr)^{\frac{J_{14,23}}{2}}
    \biggl(
        \frac{
            u \hat{q}_2 \cdot \epsilon_3 \hat{q}_{34}+s \hat{q}_{23} \hat{q}_4 \cdot \epsilon_3
        }{u \hat{q}_2 \cdot \bar{\epsilon}_3 \hat{q}_{34}+s \hat{q}_{23} \hat{q}_4 \cdot \bar{\epsilon}_3}
    \biggr)^{\frac{J_{3,4}}{2}}
    \, .
    \nn
\end{align}
Then fixing to the conformal frame and stripping off the kinematical factor, we obtain the shadow celestial amplitude
\begin{align}
    \label{eqapp: shadow amplitude}
    \wave\cA(\chi)
    &=
    2^{-\Delta_{1234}}
    \chi^{1+h_{2,1}}
    \bar{\chi}^{1+\bar{h}_{2,1}}
    (1-\chi )^{1+h_{4,123}}
    (1-\bar{\chi})^{1+\bar{h}_{4,123}}
    \\
    &\peq
    \times
    \inttt{dsdt}{\regionS}
    s^{\frac{\Delta_{34,12}}{2}}
    (-t)^{\frac{\Delta_{24,13}}{2}}
    (-u)^{\frac{\Delta_{23,14}}{2}}
    (s+t \chi )^{2 h_1-2}
    (s+t \bar{\chi})^{2 \bar{h}_1-2}
    \cT(s,t)
    \nn
    \, .
\end{align}

\subsection{OPE limit and \leading coefficient}

\textbf{\schannel.}
By the condition \eqref{eqapp: s-channel physical region}, we can safely take the OPE limit $\chi\to 0$ inside the integrals of \eqref{eqapp: shadow amplitude}, and the leading term is
\begin{equation}
    \label{eqapp: s-channel OPE behavior}
    \wave\cA(\chi)
    \sim
    \cC^{\schannel}_{\Delta_{\schannel},J_{\schannel}}(\alpha_{\schannel},\beta)
    \chi^{\hwave_{1}+h_{2}}\chib^{\hbwave_{1}+\hb_{2}}
    \, .
\end{equation}
This corresponds to an exchange operator of double-trace type.
For later convenience we introduce the parameters:
\begin{equation}
    \Delta_{\schannel}= \wave\Delta_{1}+\Delta_{2}
    \, ,
    \quad
    J_{\schannel}=\wave{J}_1+J_2
    \, ,
    \quad
    \alpha_{\schannel}= 1+\half\Delta_{23,14}
    \, ,
    \quad
    \beta= \Delta_{1234}-4
    \, .
\end{equation}
Then the \leading coefficient in \eqref{eqapp: s-channel OPE behavior} is given by
\begin{equation}
    \label{eqapp: C by scattering amplitude}
    \cC^{\schannel}_{\Delta_{\schannel},J_{\schannel}}(\alpha_{\schannel},\beta)
    =
    \inttt{dsdt}{\regionS}
    2^{-\beta-4}
    s^{\beta/2-\Delta_{\schannel}}
    (-t)^{\Delta_{\schannel}-\alpha_{\schannel}-1} (-u)^{\alpha_{\schannel}-1}
    \cT(s,t)
    \, .
\end{equation}
This is actually a two-fold Mellin transform by the change of variables from $(s,t)$ to $(\sqrt{s},-s/t)$.
Denoting the measure of inverse Mellin transform as $[d...]\eqq\frac{d...}{2\pi i}$, we obtain the relation from the \leading coefficient to the scattering amplitude
\begin{align}
    \label{eqapp: scattering amplitude by C}
    \cT(s,t)
    &=
    \intt{[d\alpha_{\schannel}][d\beta]}
    2^{\beta +3}
    s^{-\beta /2}
    \Bigl(\frac{u}{t}\Bigr)^{-\alpha_{\schannel}}
    \Bigl(-\frac{s}{t}\Bigr)^{\Delta_{\schannel}}
    \cC^{\schannel}_{\Delta_{\schannel},J_{\schannel}}(\alpha_{\schannel},\beta)
    \, .
\end{align}

\textbf{$\mathsf{t}/\mathsf{u}$-channel.}
Now we consider the \tchannel OPE. After restoring the \schannel kinematical factor and stripping off the \tchannel one, the relation between $\wave\cA$ and $\cT$ is
\begin{align}
    % \label{eq: shadow amplitude t-channel}
    \wave\cA_{\tchannel}(\chi_{\tchannel})
    &=
    (-1)^{J_{1234}}
    2^{-\Delta_{1234}}
    \chi_{\tchannel}^{1+h_{3,1}}
    \chib_{\tchannel}^{1+\bar{h}_{3,1}}
    (1-\chi_{\tchannel} )^{1+h_{4,123}}
    (1-\chib_{\tchannel})^{1+\bar{h}_{4,123}}
    \\
    &\peq \xx
    \inttt{dsdt}{\regionS}
    s^{\frac{\Delta_{34,12}}{2}}
    (-t)^{\frac{\Delta_{24,13}}{2}}
    (-u)^{\frac{\Delta_{23,14}}{2}}
    (-t-s \chi_{\tchannel} )^{2 h_1-2}
    (-t-s \chib_{\tchannel})^{2 \bar{h}_1-2}
    \cT(s,t)
    \, ,
    \nn
\end{align}
where the \tchannel cross ratio is $\chi_{\tchannel}=\chi_{\schannel}^{-1}=\frac{z_{1,3} z_{2,4}}{z_{1,2} z_{3,4}}$.

Unlike the \schannel case, here the OPE limit $\chi_{\tchannel}\to 0$ is not separated from the integration region, and there can be other types of exchange operators in this OPE channel.
Nevertheless, the exchange operator of double-trace type is still present in the \tchannel with
\begin{equation}
    \label{eqapp: tchannel leading operator}
    \Delta_{\tchannel}= \wave\Delta_{1}+\Delta_{3}
    \, ,
    \quad
    J_{\tchannel}=\wave{J}_1+J_3
    \, .
\end{equation}
Similar computation gives the \uchannel one
\begin{equation}
    \Delta_{\uchannel}= \wave\Delta_{1}+\Delta_{4}
    \, ,
    \quad
    J_{\uchannel}=\wave{J}_1+J_4
    \, .
\end{equation}
The corresponding conformal block coefficients are
\begin{align}
    \label{eqapp: C by scattering amplitude TU channel}
    &\cC^{\tchannel}_{\Delta_{\tchannel},J_{\tchannel}}
    (\alpha_{\tchannel},\beta)
    =
    (-1)^{J_{1234}}
    \inttt{dsdt}{\regionS}
    2^{-\beta-4}
    s^{\alpha_{\tchannel}-1}
    (-t)^{\beta/2-\Delta_{\tchannel}}
    (-u)^{\Delta_{\tchannel}-\alpha-1}
    \cT(s,t)
    \, ,
    \\
    &\cC^{\uchannel}_{\Delta_{\uchannel},J_{\uchannel}}
    (\alpha_{\uchannel},\beta)
    =
    (-1)^{J_{1234}}
    \inttt{dsdt}{\regionS}
    2^{-\beta-4}
    s^{\Delta_{\uchannel}-\alpha_{\uchannel}-1}
    (-t)^{\alpha_{\uchannel}-1}
    (-u)^{\beta/2-\Delta_{\uchannel}}
    \cT(s,t)
    \, ,
\end{align}
where
\begin{equation}
    \alpha_{\tchannel}= 1+\half\Delta_{34,12}
    \, ,
    \quad
    \alpha_{\uchannel}= 1+\half\Delta_{24,13}
    \, .
\end{equation}
For the \tchannel we perform the following change of variables $\chi=s/(s+t),\, \omega=\sqrt{-t}$, then
\begin{equation}
    \cC^{\tchannel}_{\Delta_{\tchannel},J_{\tchannel}}
    (\alpha_{\tchannel},\beta)
    =
    (-1)^{J_{1234}}
    \intrange{d\omega}{0}{\oo}
    \intrange{d\chi}{1}{\oo}
    2^{-\beta-3}
    \omega^{\beta -1}
    \chi^{\alpha_{\tchannel} -1}
    (\chi -1)^{-\Delta_{\tchannel}}
    \cT\LR{\frac{\omega^{2}\chi}{\chi-1},-\omega^2}
    \, .
\end{equation}
This is a two-fold Mellin transform \wrt $2^{-\beta-3}(\chi -1)^{-\Delta_{\tchannel}}\cT(\frac{\omega^{2}\chi}{\chi-1},-\omega^2)\theta(\chi-1)$,
in which the step function supports on the physical region $\theta(s+t)\theta(-t)$ for $\omega>0$. Hence the inverse relation is
\begin{equation}
    \label{eqapp: scattering amplitude by C T channel}
    \theta(s+t)\theta(-t)\cT(s,t)
    =
    (-1)^{J_{1234}}
    \intt{[d\alpha_{\tchannel}][d\beta]}
    2^{\beta +3}
    (-t)^{-\beta /2}
    \Bigl(-\frac{s}{u}\Bigr)^{-\alpha_{\tchannel}}
    \Bigl(\frac{t}{u}\Bigr)^{\Delta_{\tchannel}}
    \cC^{\tchannel}_{\Delta_{\tchannel},J_{\tchannel}}
    (\alpha_{\tchannel},\beta)
    \, .
\end{equation}
Similarly for the \uchannel, we use the change of variables $(s, t) \to (\sqrt{s+t}, -t/s)$ and obtain the inverse relation
\begin{equation}
    \label{eqapp: scattering amplitude by C U channel}
    \theta(s+t)\theta(-t)\cT(s,t)
    =
    (-1)^{J_{1234}}
    \intt{[d\alpha_{\uchannel}][d\beta]}
    2^{\beta +3}
    (-u)^{-\beta /2}
    \Bigl(-\frac{t}{s}\Bigr)^{-\alpha_{\uchannel}}
    \Bigl(-\frac{u}{s}\Bigr)^{\Delta_{\uchannel}}
    \cC^{\uchannel}_{\Delta_{\uchannel},J_{\uchannel}}
    (\alpha_{\uchannel},\beta)
    \, .
\end{equation}

\subsection{Example}

We provide an example to verify the relations \eqref{eqapp: C by scattering amplitude} and \eqref{eqapp: scattering amplitude by C}.
At tree-level, the scattering amplitude of four massless scalars with massive scalar exchange is
\begin{equation}
    \label{eq: T with massive pole}
    \cT(s,t)=
    \frac{1}{s-m^{2}+i\epsilon}+
    \frac{1}{t-m^{2}+i\epsilon}+
    \frac{1}{u-m^{2}+i\epsilon}
    \, ,
\end{equation}
and the \schannel \leading coefficient is
\begin{equation}
    \label{eq: C with massive pole}
    \cC^{\schannel}_{\Delta_{\schannel},0}(\alpha_{\schannel},\beta)
    =
    \frac{-(m^{2}-i\epsilon)^{\frac{\beta-2}{2}}\pi}{ 2^{\beta +4}\sin (\frac{\pi  \beta }{2})}
    \Big(
        e^{\frac{1}{2} i \pi \beta }
        B(\gamma -\alpha_{\schannel} ,\alpha_{\schannel} )
        +
        B(-\alpha_{\schannel} -\frac{\beta }{2}+\gamma ,\alpha_{\schannel} )
        +
        B(\gamma -\alpha_{\schannel} ,\alpha_{\schannel} -\frac{\beta }{2})
    \Big)
    \, .
\end{equation}
In the physical region $I$, only the first term in \eqref{eq: T with massive pole} can be put on-shell, and the phase factor $e^{\frac{1}{2} i \pi \beta }$ in \eqref{eq: C with massive pole} is due to $\arg(-m^2+i \epsilon)=\pi-\epsilon$.

%%%%%%%%%%%%%%%%%%%%%%%%%%%%%%%%%%%%%%%%%%%%%%%%%%%%%%%%%%%%%%%%

\section{Three-point celestial amplitudes of gluons}\label{app: three-point gluon}

In this section we use a regularization method to compute the celestial amplitude of three gluons $1^{-}+2^{-}\to3^{+}$. We confirm the result by another independent method in the upcoming paper \cite{celestialOPE}.

The momentum conservation $\deltaMC(q_{1}+q_{2}-q_{3})$ is singular and contains three different components:
\begin{align}
    \text{colinear:} & \quad
    z_1 = z_2 = z_3
    \, , \quad
    \bar{z}_1 = \bar{z}_2 = \bar{z}_3
    \, , \quad
    \omega_1 + \omega_2 = \omega_3
    \, ,
    \\
    \text{1-soft:} & \quad
    z_2 = z_3
    \, ,\quad
    \bar{z}_2 = \bar{z}_3
    \, , \quad
    \omega_2 = \omega_3
    \, , \quad
    \omega_1 = 0
    \, ,
    \\
    \text{2-soft:} & \quad
    z_1 = z_3
    \, ,\quad
    \bar{z}_1 = \bar{z}_3
    \, , \quad
    \omega_1 = \omega_3
    \, , \quad
    \omega_2 =0
    \, .
\end{align}
To resolve the singularity we will introduce an infinitesimal regulator $\varepsilon>0$, and consequently the Ward identity of the scattering amplitude gets broken. Hence it is necessary to use the full conformal wavefunction to obtain a conformal covariant celestial amplitude.

We adopt the shadow basis for the 3-rd particle, then the sign of conformal spin $J_{3}$ will be flipped.
For the massless gluon with helicity $a=\pm 1$, the full conformal wavefunction integrated with a scattering amplitude is
\begin{align}
    \intt{d\omega} \omega^{\Delta-1}
    \left(
        (\pp_{a}q^{\mu}) \cT_{\mu}(q(\omega,z),\cdots)
        +
        \frac{1}{\Delta-1}\pp_{a} \left(\qhat^{\mu}\cT_{\mu}(q(\omega,z),\cdots)\right)
    \right)
    \, .
\end{align}
Then the celestial amplitude is given by
\begin{align}
    \cA
    &=
    \intrange{d\omega_1}{0}{\oo}
    \omega_1^{\Delta_1-1}
    \left(
        \bar{\epsilon}_1^\mu+
        \frac{\pp_{\bar{z}_1}\cdot \hat{q}_1^\mu}{\Delta_1-1}
    \right)
    \intrange{d\omega_2}{0}{\oo}
    \omega_2^{\Delta_2-1}
    \left(
        \bar{\epsilon}_2^\nu+
        \frac{\pp_{\bar{z}_2}\cdot \hat{q}_2^\nu}{\Delta_2-1}
    \right)
    \\
    &\peq \xx
    \int d^{4}q_{3'}\delta(q_{3'}^{2})
    2^{\Delta_{3}-1}
    \left(-q_{3'} \cdot \hat{q}_3\right)^{-\Delta_3}
    \left(g^{\rho \rho'}-\frac{\hat{q}_3^{\rho'} \hat{q}_{3'}^\rho}{\hat{q}_{33'}}-\frac{\hat{q}_3^\rho \hat{q}_{3'}^{\rho'}}{\hat{q}_{33'}}\right)
    \bar{\epsilon}_{3,\rho'}
    \nn
    \\
    &\peq \xx
    \left(g_{\nu\rho} (q_{3',\mu}+q_{2,\mu})+g_{\mu\rho} (-q_{3',\nu}-q_{1,\nu})+(q_{1,\rho}-q_{2,\rho}) g_{\mu\nu}\right)
    \delta^{(4)}(q_1+q_2-q_{3'})
    \nn
    \, ,
\end{align}
where $\qhat_{ij}\eqq \qhat_{i}\cdot\qhat_{j}$ and the derivatives
$\pp_{\bar{z}_1}\cdot,\, \pp_{\bar{z}_2}\cdot$ act on all the rest of expressions.
We first perform the $q_{3'}$-integral by the momentum conservation, leading to four terms
\begin{equation}
    \cA=
    \cA_{0}+
    \pp_{\zb_{1}}\cA_{1}+
    \pp_{\zb_{2}}\cA_{2}+
    \pp_{\zb_{1}}\pp_{\zb_{2}}\cA_{3}
    \, .
\end{equation}
where each one is of the form
\begin{equation}
    \cA_{i}
    \sim
    \intrange{d\omega_1}{0}{\oo}
    \intrange{d\omega_2}{0}{\oo}
    \omega_1^{\Delta_1+n_{1}}
    \omega_2^{\Delta_2+n_{2}}
    \delta (2 \omega_1 \omega_2 \hat{q}_{12})
    (-\omega_1 \hat{q}_{13}-\omega_2 \hat{q}_{23})^{-\Delta_3-1}
    \xx (\cdots)
\end{equation}
for some integer $n_{1},\, n_{2}$.

As discussed before, we introduce a infinitesimal regulator $\varepsilon>0$ to $\delta (2 \omega_1 \omega_2 \hat{q}_{12}+\varepsilon)$.
Then the $\omega_{2}$-integral can be preformed by the $\delta$-function, and the remaining $\omega_{1}$-integral is of the form
\begin{equation}
    \intrange{d\omega_1}{0}{\oo}
    (a \omega_{1}^{2}+b)^{c}\omega_{1}^{d}
    =
    \frac{1}{2} a^{\frac{-1-d}{2}}
    b^{\frac{1+2 c+d}{2}}
    \frac{
        \gm{\frac{-1-2 c-d}{2}}
        \gm{\frac{1+d}{2}}
    }{\gm{-c}}
    \, .
\end{equation}
The resulting $\cA$ takes the form of the standard conformal correlator with the three-point coefficient being
\begin{align}
    \label{eq: AASA}
    \mathscr{C}^{(\Delta_{1},-)(\Delta_{2},-)(\Delta_{3},-)}_{AA\to\wave{A}}
    =
    \lim_{\varepsilon\to 0^{+}}
    2^{\Delta_{3,12}-2}
    \varepsilon^{\frac{\Delta_{12,3}-1}{2}}
    \frac{(\Delta_{123}-3) (\Delta_{123}-1)}{(\Delta_1-1) (\Delta_2-1)}
    \frac{
        \gm{\frac{1+\Delta_{13,2}}{2}}
        \gm{\frac{1+\Delta_{23,1}}{2}}
    }{\gm{1+\Delta_3}}
    \, .
\end{align}

We notice that when $\Delta_{3}=\Delta_{12}-1$, the three-point coefficient reduces to $\half B(\Delta_{1}-1,\Delta_{2}-1)$, proportional to the coefficient in the collinear OPE.
We will explore more on the compatibility between three-point coefficient and OPE coefficient in \cite{celestialOPE}.

The three-point coefficient $\mathscr{C}_{\wave{A}A\to A}$ can be obtained by performing shadow transforms on \eqref{eq: AASA}, given by
\begin{equation}
    \mathscr{C}^{(\Delta_{1},+)(\Delta_{2},-)(\Delta_{3},+)}_{\wave{A}A\to A}
    =
    \lim_{\varepsilon\to 0^{+}}
    \frac{-2^{\Delta_{1,23}}\varepsilon^{\frac{\Delta_{23,1}-1}{2}}}{\Delta_2-1}
    \frac{
        \gm{-\Delta_1}
        \gm{\frac{1+\Delta_{12,3}}{2}}
        \gm{\frac{-1+\Delta_{123}}{2}}
        \gm{\frac{1+\Delta_{13,2}}{2}}
    }{\gm{\Delta_1} \gm{\Delta_3} \gm{\frac{-1+\Delta_{2,13}}{2}} }
    \, .
\end{equation}
Under the condition $\Delta_{1}=2-\Delta_{2},\, \Delta_{3}=1$, the three-point coefficient including one primary descendant is
\begin{equation}
    \mathscr{C}^{(2-\Delta_{1},+)(\Delta_{1},-)(2,0)}_{\wave{A}A\to\ppb A}
    =
    \lim_{\varepsilon\to 0^{+}}
    \frac{
        2^{1-2 \Delta_1} \Delta_{1}
    }{1-\Delta_1}
    \varepsilon^{\Delta_1-1}
    \, .
\end{equation}
Now as a distribution of $\Delta_{1}$, the coefficient should be understood as
\begin{equation}
    \lim_{\varepsilon\to 0^{+}}
    \intrange{\frac{d\Delta_{1}}{2\pi i}}{1-i\oo}{1+i\oo}
    \frac{
        2^{1-2 \Delta_1} \Delta_{1}
    }{1-\Delta_1}
    \varepsilon^{\Delta_1-1}
    f(\Delta_{1})
    \, ,
\end{equation}
where $f(\Delta_{1})$ is a test function decaying sufficient fast at infinity. The pole $\Delta_{1}=1$ lies on the contour, hence the integral should be understood as the principal value and provides an extra factor $\half$, leading to
\begin{equation}
    \mathscr{C}^{(2-\Delta_{1},+)(\Delta_{1},-)(2,0)}_{\wave{A}A\ppb A}
    =
    \frac{1}{4}\delta_{\CC}(\Delta_{1}-1)
    \, .
\end{equation}
%

%%%%%%%%%%%%%%%%%%%%%%%%%%%%%%%%%%%%%%%%%%%%%%%%%%%%%%%%%%%%%%%%

\section{Reconstruction details}\label{app: reconstruction}

In this section we first provide in detail the \tchannel reconstruction of the graviton MHV amplitude, then sketch the reconstruction of the MHV amplitudes in the pure Yang-Mills theory and the Einstein-Yang-Mills theory.
We denote an incoming/outgoing gluon operator with color index $a$, conformal dimension $\Delta$ and spin $\pm1$ by $\gluon^{a,\inn/\out}_{\Delta,+/-}$, and a graviton one with spin $\pm2$ by $\graviton^{\inn/\out}_{\Delta,++/--}$.

\subsection{Graviton MHV amplitude from \mathInTitle{\mathsf{t}}-channel OPE}

We consider the graviton MHV scattering $1^{--}+2^{--}\to 3^{++}+4^{++}$. In the \tchannel OPE, by \eqref{eqapp: tchannel leading operator} the conformal weights of the \leading operator $\cO^{\tchannel}$ are $\Delta_{\tchannel}=\Delta_{3,1}+2\, , J_{\tchannel}=4$.
There are four graviton operators in the OPE
\begin{align}
    \label{eqapp: graviton tchannel OPE}
    &
    \graviton_{\Delta_{2},--}^{\inn}(z_{2})
    \graviton_{\Delta_{4},++}^{\out}(z_{4})
    \\
    &
    \sim
    \frac{\zb_{2,4}}{z_{2,4}}\left(
        B(\Delta_{2}+3,-1-\Delta_{24})
        \graviton_{\Delta_{24},--}^{\out}(z_{4})
        +
        B(\Delta_{4}-1,-1-\Delta_{24})
        \graviton_{\Delta_{24},--}^{\inn}(z_{4})
    \right)
    \nn
    \\
    &
    +
    \frac{z_{2,4}}{\zb_{2,4}}\left(
        B(\Delta_{2}-1,-1-\Delta_{24})
        \graviton_{\Delta_{24},++}^{\out}(z_{4})
        +
        B(\Delta_{4}+3,-1-\Delta_{24})
        \graviton_{\Delta_{24},++}^{\inn}(z_{4})
    \right)
    \, .
    \nn
\end{align}

As in the main text, we assume the minimality condition, \ie there are no other contributions except for the primary descendants of the gravitons to the \leading operator $\cO^{\tchannel}$.
To match the exchange gravitons with positive spin in \eqref{eqapp: graviton tchannel OPE}, we have
\begin{equation}
    \label{eqapp: graviton tchannel leading operator}
    \cO^{\tchannel}_{\Delta_{3,1}+2,4}
    =
    c_{1}
    \pp^{2}\graviton_{\Delta_{24},++}^{\out}
    +
    c_{2}
    \pp^{2}\graviton_{\Delta_{24},++}^{\inn}
    \, ,
\end{equation}
then the conformal dimension matching gives
\begin{equation}
    \label{eqapp: graviton tchannel Delta matching}
    2-\Delta_1+\Delta_3=\Delta_2+\Delta_4+2=-1+\varepsilon
    \, .
\end{equation}
Here we have introduced an infinitesimal regulator $\varepsilon$ to tame the divergence in the OPE coefficients later. We will take the limit $\varepsilon\to 0$ at the end of the calculation to restore the conformal covariance ensured by the primary descendant condition.

Under this condition, the gravitons with negative spin in \eqref{eqapp: graviton tchannel OPE} does not contribute to $\cO^{\tchannel}$ since the corresponding primary descendants are $\pp^{6}\graviton_{\Delta_{24},--}^{\inn/\out}$ and then the conformal dimension matching $\Delta_{3,1}+2=\Delta_{24}+6=3$ contradicts with \eqref{eqapp: graviton tchannel Delta matching}.

Similar to the \schannel, with \eqref{eqapp: graviton tchannel Delta matching} the three-point coefficient of $\cO^{\tchannel}$ can be fixed as $\deltaComplex{2\Delta_3-2}$ due to the scaling behavior $(\Delta_{3}+3-\varepsilon)+\Delta_{3}+(-3+\varepsilon)=2$.
The OPE coefficients can be derived from \eqref{eqapp: graviton tchannel OPE}, leading to
\begin{align}
    \label{eqapp: tchannel Cs Graviton}
    \cC^{\tchannel}
    =
    \frac{1}{2}c_{1} B(\Delta_{2}+1,-3-\Delta_{24})\deltaComplex{2\Delta_3-2}
    +
    \frac{1}{2}c_{2} B(\Delta_{4}+3,-3-\Delta_{24})\deltaComplex{2\Delta_3-2}
    \, ,
\end{align}
where we have absorbed the numerical prefactors of three-point coefficients into $c_{1}$ and $c_{2}$.
From \eqref{eqapp: graviton tchannel Delta matching} we have $\Delta_{\tchannel}=-1+\varepsilon$, $\alpha_{\tchannel}=\Delta_{4}+1$ and $\beta=2\Delta_{3}-4$, then \eqref{eqapp: tchannel Cs Graviton} can be rewritten as
\begin{equation}
    \label{eqapp: tchannel Cs Graviton 2}
    \cC^{\tchannel}
    =
    \frac{1}{2}
    \deltaComplex{\beta+2}
    \bigl(
        c_{1} B(\varepsilon-\alpha-1,-\varepsilon)
        +
        c_{2} B(\alpha+2,-\varepsilon)
    \bigr)
    \, .
\end{equation}

Inserting \eqref{eqapp: tchannel Cs Graviton 2} into \eqref{eqapp: scattering amplitude by C T channel} we obtain the stripped amplitude
\begin{align}
    \label{eq: cT from cC tchannel}
    &\theta(s+t)\theta(-t)\cT_{\graviton}(s,t)
    =
    \intt{[d\alpha_{\tchannel}][d\beta]}
    2^{\beta +2}
    (-t)^{-\beta /2}
    \Bigl(-\frac{s}{u}\Bigr)^{-\alpha_{\tchannel}}
    \Bigl(\frac{t}{u}\Bigr)^{-1+\varepsilon}
    \\
    &\quad \xx
    \deltaComplex{\beta+2}
    \bigl(
        c_{1} B(\varepsilon-\alpha-1,-\varepsilon)
        +
        c_{2} B(\alpha+2,-\varepsilon)
    \bigr)
    \, .
    \nn
\end{align}
If the $\alpha_{\tchannel}$-contour lies in $\Re \alpha_{\tchannel}>-2$ and is deformed to the right half-plane then
\begin{equation}
    \label{eqapp: cT Graviton t-channel 1}
    \cT_{\graviton}(s,t)
    =
    c_{1}
    s^{2-n_1} t^{-1}(-u)^{n_1}
    \, ,
    \textInMath{for}
    n_1\in\NN
    \, ,
\end{equation}
while if lies in $\Re \alpha_{\tchannel}<-2$ then
\begin{equation}
    \label{eqapp: cT Graviton t-channel 2}
    \cT_{\graviton}(s,t)
    =
    c_{2} s^{3+n_2} t^{-1}(-u)^{-1-n_2}
    -(c_{1}+c_{2})s^{2}t^{-1}
    \, ,
    \textInMath{for}
    n_2\in\NN
    \, .
\end{equation}

In the main text, we have reconstructed from the \schannel OPE a family of scattering amplitudes
\begin{equation}
    \label{eqapp: cT Graviton s-channel}
    \cT_{\graviton}(s,t)
    =
    \threeptN_{\graviton}
    (-1)^{n_{\schannel}}s^{3} (-t)^{n_{\schannel}-2} (-u)^{-n_{\schannel}}
    \, ,
    \textInMath{for}
    n_{\schannel}\in\NN
    \, .
\end{equation}
Comparing with \eqref{eqapp: cT Graviton t-channel 1} and \eqref{eqapp: cT Graviton t-channel 2} we obtain $n_{\schannel}=1$, $n_{2}=0$ and $\threeptN_{\graviton}=-c_{1}=c_{2}$.

There is a technical issue to be pointed out.
For \eqref{eq: cT from cC tchannel}, if deforming the $\alpha_{\tchannel}$-contour to the left half-plane and picking up the poles of $\gm{\alpha_{\tchannel}+2}$, this requires $-s/u<1\implies t>0$ and leads to a nonvanishing result, contradicting to the step function $\theta(-t)$.
We leave this in the future work.

\subsection{MHV amplitude in Einstein-Yang-Mills theory}

We consider the scattering $1^{-}+2^{-}\to 3^{++}+4^{+}$ in Einstein-Yang-Mills theory.
Since the details of reconstruction have been discussed in the pure gravity theory, we will only outline the key steps.

In the \schannel, the OPE is given by
\begin{align}
    \graviton_{\Delta_{3},++}^{\out}(z_{3})
    \gluon_{\Delta_{4},+}^{a,\out}(z_{4})
    \sim
    \frac{\zb_{3,4}}{z_{3,4}}
    B(\Delta_{3}-1,\Delta_{4})
    \gluon_{\Delta_{24},+}^{a,\out}(z_{4})
    \, .
\end{align}
The \leading operator is a colored scalar, and assuming the minimality condition we have $\cO^{a,\schannel}_{\Delta_{2,1}+2,0}=\bar{\partial}\gluon^{a,\out}_{\Delta_{34},+}$. Then the conformal dimension matching is
\begin{align}
    2-\Delta_1+\Delta_2=\Delta_3+\Delta_4+1=2
    \, ,
\end{align}
and the \leading coefficient is
\begin{align}\label{eq:Cs graviton-gluon}
    \mathcal{C}^{\schannel}
    =
    \half c_{\schannel}\delta(\Delta_2-1)B(\Delta_3,\Delta_4)
    \, ,
\end{align}
for some numerical constant $c_{\schannel}$.
Substituting \eqref{eq:Cs graviton-gluon} into \eqref{eqapp: C by scattering amplitude}, we obtain
\begin{align}
    \label{eqapp: gghg}
    \cT(s,t)=
    4
    (-1)^{n_{\schannel}}
    c_{\schannel}
    s^{3/2}
    (-t)^{n_\schannel-\frac{1}{2}}
    (-u)^{-n_\schannel-\frac{1}{2}}
    \, ,
    \textInMath{for}
    n_{\schannel}\in\NN
    \, .
\end{align}

In the \tchannel, the OPE is given by
\begin{align}
    &
    \gluon_{\Delta_{2},-}^{a,\inn}(z_{2})
    \gluon_{\Delta_{4},+}^{b,\out}(z_{4})
    \\
    &
    \sim
    \frac{f^{abc}}{z_{2,4}}\left(
        -B(\Delta_{2}+1,1-\Delta_{24})
        \gluon_{\Delta_{24}-1,-}^{c,\out}(z_{4})
        +
        B(\Delta_{4}-1,1-\Delta_{24})
        \gluon_{\Delta_{24}-1,-}^{c,\inn}(z_{4})
    \right)
    \nn
    \\
    &+
    \frac{f^{abc}}{\zb_{2,4}}\left(
        B(\Delta_{2}-1,1-\Delta_{24})
        \gluon_{\Delta_{24}-1,+}^{c,\out}(z_{4})
        % +
        -
        B(\Delta_{4}+1,1-\Delta_{24})
        \gluon_{\Delta_{24}-1,+}^{c,\inn}(z_{4})
    \right)
    \, .
    \nn
\end{align}
Assuming the minimality condition, the \leading operator is
\begin{equation}
    \cO^{a,\tchannel}_{\Delta_{3,1}+2,4}
    =
    c_{1}
    \pp^{2}\gluon_{\Delta_{24}-1,+}^{\out}
    +
    c_{2}
    \pp^{2}\gluon_{\Delta_{24}-1,+}^{\inn}
    \, ,
\end{equation}
and the conformal dimension matching is
\begin{align}
    2-\Delta_1+\Delta_3=\Delta_2+\Delta_4+1=\varepsilon
    \, .
\end{align}
Similar to the pure gravity case, according to the position of $\alpha_{\tchannel}$-contour, the stripped amplitude is either
\begin{align}
    \cT(s,t)=
    4
    (-1)^{n_{1}}
    s^{3/2}
    (-t)^{n_1-\frac{1}{2}}
    (-u)^{-n_1-\frac{1}{2}}
    \, ,
    \textInMath{for}
    n_{1}\in\NN
    \, .
\end{align}
or
\begin{align}
    \cT(s,t)
    &=
    2 c_2
    s^{\frac{3}{2}+n_2}
    (-u)^{-\frac{1}{2}-n_2}
    (-t)^{-\frac{1}{2}}
    -2 c_2
    s^{3/2}
    (-t)^{-\frac{1}{2}}
    (-u)^{-\frac{1}{2}}
    \nn
    \\
    &\peq
    -2(u c_1+t c_2)
    s^\frac{1}{2}
    (-t)^{-\frac{1}{2}}
    (-u)^{-\frac{1}{2}}
    \, ,
    \textInMath{for}
    n_{2}\in\NN
    \, .
\end{align}
Comparing with \eqref{eqapp: gghg} we find that there is only one solution: $n_{\schannel}=n_{2}=0$ and $c_{1}=c_{2}=2c_{\schannel}$, then
\begin{align}
    \cT^{abd}(1^{-}2^{-}3^{++}4^{+})
    =
    f^{abd}
    s^{3/2}(-t)^{-\frac{1}{2}}(-u)^{-\frac{1}{2}}
    \frac{z_{1,2} z_{1,4}^{1/2} \bar{z}_{1,3}^{1/2} \bar{z}_{3,4}^{3/2}}{\bar{z}_{1,2} \bar{z}_{1,4}^{1/2} z_{1,3}^{1/2} z_{3,4}^{3/2}}
    \delta^{(4)}(q_1+q_2-q_3-q_4)
    \, .
\end{align}

\subsection{Gluon MHV amplitude}
We consider the scattering process $1^-+2^-\rightarrow 3^++4^+$ in the pure Yang-Mills theory.
In the previous cases, we apply the CFT crossing symmetry to fix the contour ambiguity. 
Here due to the presence of color factor, we also need the permutation symmetry of identical particles. The corresponding stripped amplitude takes the form
\begin{align}
    \cT^{abcd}(s,t)
    &=
    f^{abe}f^{cde}
    \cT_{1}(s,t)
    +
    f^{ace}f^{bde}
    \cT_{2}(s,t)
    =
    f^{ade}f^{bce}
    \cT_{3}(s,t)
    +
    f^{ace}f^{bde}
    \cT_{4}(s,t)
    \, .
\end{align}
Using the Jacobi identity of structure constants, we find that
\begin{align}\label{eq:Jacobi}
    \cT_1(s,t)=-\cT_3(s,t)
    \, ,\quad 
    \cT_2(s,t)=\cT_3(s,t)+\cT_4(s,t)
    \, .
\end{align}
The permutation symmetry $(34)$ of the full amplitude $\cT^{abcd}(1^{-}2^{-}3^{+}4^{+})$ leads to
\begin{align}\label{eq:permutation}
    \cT_1(s,u)+\cT_1(s,t)=-\cT_2(s,u)
    \, ,\quad  
    \cT_{2}(s,t)=\cT_2(s,u)\, ,\quad 
    \cT_3(s,t)=\cT_4(s,u)
    \, ,
\end{align}
where we have used the fact that the little group factor $H$ is invariant under the permutation $(34)$.

According to the color structures, 
$\cT_1$ can be determined by $\ppb\gluon_{\Delta,+}$ in the \schannel OPE, 
$\cT_2$, $\cT_4$ can be determined by $\pp\gluon_{\Delta,+}$ in the \tchannel, 
and $\cT_3$ can be determined by $\pp\gluon_{\Delta,+}$ in the \uchannel. 
Like in the previous cases, $\cT_i$-s can only be fixed up to some numerical constants by OPE, and the relative coefficients can be determined by \eqref{eq:Jacobi} and \eqref{eq:permutation}, leading to
\begin{align}
    \cT^{abcd}(1^{-}2^{-}3^{+}4^{+})=\left(f^{abe}f^{cde}\frac{s}{u}+f^{ace}f^{bde}\frac{s^2}{tu}\right)
    \frac{z_{1,2} \bar{z}_{3,4}}{z_{3,4} \bar{z}_{1,2}}
    \delta^{(4)}(q_1+q_2-q_3-q_4)
    \, .
\end{align}
%

%%%%%%%%%%%%%%%%%%%%%%%%%%%%%%%%%%%%%%%%%%%%%%%%%%%%%%%%%%%%%%%%

\newpage
\printbibliography

%%%%%%%%%%%%%%%%%%%%%%%%%%%%%%%%%%%%%%%%%%%%%%%%%%%%%%%%%%%%%%%%

\end{document}